\documentclass[preprint,12pt]{elsarticle}

\usepackage{amssymb}
\usepackage{amsmath}
\usepackage{array}
\usepackage{epstopdf}

\journal{Optics Communications}

\begin{document}

\begin{frontmatter}



\title{Dispersion blue-shift in an aperiodic Bragg reflection waveguide}

\author[rian]{Volodymyr~I.~Fesenko\corref{correspond}}
\ead{v.i.fesenko@ieee.org}
\author[rian]{Vladimir~R.~Tuz}

\address[rian]{Institute of Radio Astronomy of National Academy of Sciences of Ukraine, Kharkiv, Ukraine}
\cortext[correspond]{Institute of Radio Astronomy of National
Academy of Sciences of Ukraine, 4, Chervonopraporna St., Kharkiv
61002, Ukraine}

\begin{abstract}
A particular feature of an aperiodic design of cladding of Bragg
reflection waveguides to demonstrate a dispersion blue-shift is
elucidated. It is made on the basis of a comparative study of
dispersion characteristics of both periodic and aperiodic
configurations of Bragg mirrors in the waveguide system, wherein for
the aperiodic configuration three procedures for layers alternating,
namely Fibonacci, Thue--Morse and Kolakoski substitutional rules
are considered. It was found out that, in a Bragg reflection
waveguide with any considered aperiodic cladding, dispersion curves
of guided modes appear to be shifted to shorter wavelengths compared
to the periodic configuration regardless of the modes polarization.
\end{abstract}


\end{frontmatter}



\section{Introduction}
\label{sec:intro}

As is known,  in contrast with conventional optical waveguides
based on the effect of the total internal reflection inside a
high-index core \cite{snyder1983}, distinctive features of Bragg
reflection waveguides are influenced by a multilayered configuration
of their composite cladding \cite{yeh_OptCommun_1976}, which has a
form of a stack of alternating high- and low-index layers. Indeed,
such a composition of cladding leads to formation of photonic
bandgaps in the spectra of the multilayered system resulting in
light confinement within a  low-index core which is usually
considered to be an air gap. Such photonic bandgap guidance brings
several attractive features to the waveguide characteristics
\cite{Johnson_OE_01, russel_Science_2003}, in particular, since most
of light is guided inside a low-index core, losses and nonlinear
effects can be significantly suppressed compared to the conventional
high-index guiding waveguides.

Furthermore, even in the simplest symmetric configuration of Bragg
reflection waveguides there is a set of unique optical features that
are unattainable in conventional waveguides since the former ones
are highly dispersive due to their complicate geometry (rather than
just to their material composition). As such features we can mention
that in Bragg reflection waveguides \cite{west_JOSAB_2006,
li_josa_2007}: (i) each guided mode has several cutoff points, which
results in ability to design a waveguide supported only the
high-order modes instead of the fundamental one; (ii) there is a
possibility to lose a specific mode due to shrinking of the photonic
bandgap into point; (iii) some specific modes with negative order
can appear when a system consists of a thin enough guiding layer. On
the other hand, Bragg reflection waveguides exhibit an extra degree
of freedom for optimizing their optical characteristics through
utilization of particular designs of the cladding, among which an
asymmetric mirrors design \cite{li_OptCommun_2008,
li_OptExpress_2009}, layers chirping in the cladding
\cite{nistad_OptCommun_2006, Pal_OptQuantElectron_2007}, and placing
the matched layers between the core and cladding
\cite{Abolghasem_QE_2009} may be referred.

In this paper we report on the other possibility of tuning the operation
bandwidth of Bragg reflection waveguides by applying an another
special design to the multilayered cladding. Our motivation in this
study is in follows. If we have a look from the viewpoint  of
practical applications, both the number of layers in the cladding
and their thicknesses should be as small as possible. It is due to
the fact that the production of waveguides with thinner layers requires
fewer amount of materials for deposition, and thus less time for the
process. At the same time, it contradicts with the technological limits
that are imposed on the possibility of producing layers with a very small
thicknesses. Besides, it is difficult to reach precise step-index profiles
with sharp boundaries, and this problem is more significant for
structures with the thinnest layers. In this context, it is of
interest to find such a configuration of layers in the cladding,
which would allow one to shift the operation frequency of the
waveguide into the higher band without changing number of layers and
their thicknesses.

It is obvious that through the solution of an optimization problem
\cite{I_jlwt2004} it is possible to find the best spatial
distribution of layers in a random sequence that produces a maximal
shift, however, in this consideration we prefer to deal with
deterministically ordered structures instead of completely random
ones, since in the first case we are able to definitely predict the
spectral properties of the system \cite{Macia_RPP_06,
Shramkova_PhB_2011, fesenko_SPIE_2014, Fesenko_NATO_2014,
fesenko_PIERM_2015}, i.e. the width and position of stopbands on the
frequency scale, which are key characteristics that are taken into consideration
when designing the Bragg reflection waveguides. As noted in our previous papers
\cite{Tuz_JOSA_09, Tuz_WRCM_09, Tuz_JModOpt_10, fesenko_WRCM_2014},
stopbands in the spectra of aperiodic structures can appear to be
shifted in comparison with the stopbands in the spectra of periodic
structures assuming the same material parameters and number of
layers in the multilayered systems. Therefore, in this paper, in
order to provide a comparative study, we consider spectral features
and dispersion characteristics of a Bragg reflection waveguide
having either periodic or aperiodic configuration of layers in the
cladding. For an aperiodic configuration we have a choice between
three well known aperiodic chains produced through substitution
rules of Fibonacci \cite{Kohmoto_PhysRevLett_1987}, Thue--Morse
\cite{Liu_PhysRevB_1997}, and Kolakoski \cite{Sing04} sequences.

\section{Theoretical Description}
\label{sec:problem}

\begin{figure}[ht]
\begin{center}
\begin{tabular}{c}
\includegraphics[height=4.0cm]{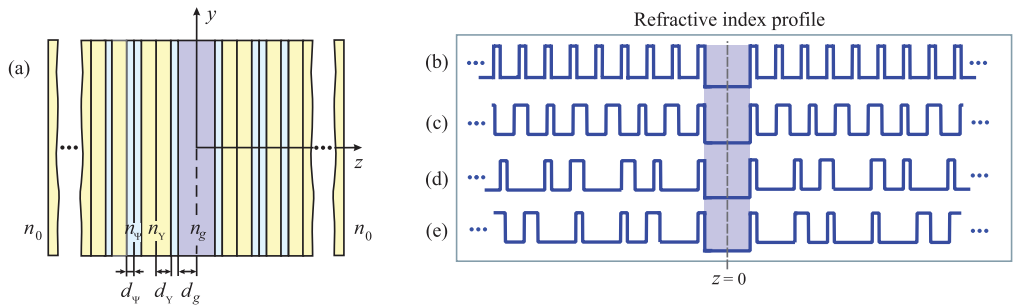}
\end{tabular}
\end{center}
\caption[example] {\label{fig:fig_1} (a) The schematic of a
symmetrical Bragg reflection waveguide that consists of a core layer
sandwiched between two aperiodic mirrors. Index profiles of the
waveguide's cladding where layers are arranged according to the
generation rules of (b)~periodic, (c)~Fibonacci, (d)~Thue--Morse,
and (e)~Kolakoski sequences.}
\end{figure}

We consider a Bragg reflection waveguide (Fig.~\ref{fig:fig_1}a)
that is made of a low-index  core layer (in particular, an air gap)
sandwiched between two identical either periodic or aperiodic
one-dimensional Bragg mirrors formed by stacking together layers of
two different sorts $\Psi$ and $\Upsilon$, which have thicknesses
$d_\Psi$, $d_\Upsilon$ and refractive indices $n_\Psi$ and
$n_\Upsilon$, respectively. The numbers of constitutive layers of
each sort are defined as $N_\Psi$ and $N_\Upsilon$. The structure
inhomogeneity (i.e. the variation of the refractive index) extends along
the $z$-axis, and in this direction the system is finite, i.e. we
suppose that mirrors on either side of the core layer consist of a
finite number $N$ of the constitutive layers. In other two
directions $x$ and $y$ the structure is invariant and infinite. In such
a geometry the axis of symmetry of the structure under study
corresponds to the middle of the core layer which is at the line $z
= 0$, where the core layer has thickness $2d_g$ and refractive index
$n_g$.

Therefore, we study a Bragg reflection
waveguide having a core layer with thickness $2d_g =
\lambda_{qw}/2n_g$, where
$\lambda_{qw}=1\mu\mathrm m$. As constituents for the cladding
composition we utilize a combination of layers made of GaAs and
oxidized AlAs, whose refractive indices at the given wavelength
$\lambda_{qw}$ are $n_\Psi = 3.50$ and $n_\Upsilon = 1.56$,
respectively. The thicknesses $d_\Psi$ and $d_\Upsilon$ are chosen to be $74.5$~nm for
GaAs layers and $208.2$~nm for AlAs-oxide layers in order to provide the
operation bandwidth to be within the first telecommunication window.

As an aperiodic configuration three alternative designs of the
waveguide's mirrors are investigated. Thus, a comparative study
between the waveguides with the multilayered cladding altered
according to the substitutional rules of Fibonacci
\cite{Kohmoto_PhysRevLett_1987}, Thue--Morse
\cite{Liu_PhysRevB_1997}, and Kolakoski \cite{Sing04} sequences is
provided (see, \ref{sec:ApA}).

In the chosen structure configuration, each guided mode of TE
polarization $\{E_x, H_y, H_z\}$ or TM polarization
$\{H_x,E_y,E_z\}$ propagates along the $y$-axis with its own propagation
constant $\beta$. As the mirrors on each side of the waveguide core
layer are the same (i.e. the Bragg reflection waveguide is
symmetrical about the $z$-axis, $n(-z) = n(z)$, as it is depicted in
Fig.~\ref{fig:fig_1}~b-e), the equations for waves travelling back
and forth inside the channel regardless of the type of polarization
can be joined on the boundaries $z = d_g$ and $z=-d_g$ into the next
system
\begin{equation}
      \label{eq:dispersionsyst}
\left\{
\begin{aligned}
&a_0\exp(-ik_{zg}d_g)=Rb_0\exp(ik_{zg}d_g),\\
&b_0\exp(-ik_{zg}d_g)=Ra_0\exp(ik_{zg}d_g),
\end{aligned}
\right.
\end{equation}
from which the relation between amplitudes can be found
\begin{equation}
      \label{eq:dispersionamp}
b_0=a_0R\exp(2ik_{zg}d_g).
\end{equation}
Here $k_{zg}=k_0(n_g^2-n_{eff}^2)^{1/2}$ is the transverse
wavenumber in the core, $n_{eff}=\beta/k_0$ is introduced as an
effective refractive index for each particular guided mode,
$k_0=\omega/c$ is the free space wavenumber, and $R$ is the complex
reflection coefficient of the Bragg mirror which is depended on the
wave polarization. The reflection coefficient $R$ can be derived
engaging the transfer matrix formalism \cite{Tuz_JOSA_09,
Tuz_WRCM_09, Tuz_JModOpt_10} (see, \ref{sec:ApB}).

Eliminating amplitudes $a_0$ and $b_0$ from system
(\ref{eq:dispersionsyst}), the dispersion equation for the guided
modes of the Bragg reflection waveguide is obtained as

\begin{equation}
      \label{eq:dispersion}
1-R^2\exp\left [4ik_0d_g\sqrt{n_g^2-n_{eff}^2}~\right]=0.
\end{equation}
This equation is further solved numerically to find out a function
of the propagation constant $\beta$ versus frequency $\omega$. The
resulting propagation constant $\beta$ is sought in the field of complex numbers due to
the existence of intrinsic losses in the waveguide constitutive materials and
energy leakage through the outermost layers because the number
of layers $N$ in the Bragg mirrors is a finite quantity.

The proper selection of the constitutive materials suggests the preference of substances with low internal losses.
According to \cite{Adachi_JApplPhys_1985} the imaginary parts of the dielectric constants of the chosen GaAs and oxidized AlAs
 are negligibly small within the wavelength region of interest. Therefore, the main source of losses in the Bragg reflection waveguides under consideration
is the energy leakage through their finite cladding. These losses can be estimated via the expression \cite{west_JOSAB_2006}:

\begin{equation}
\label{eq:Loss}
\text{Loss[dB/cm]} = \frac{-\lambda\ln|R|}{40 n_g d_g^2 \sqrt{ 1- \left(\lambda/4n_gd_g\right)^2}}.
\end{equation}

\begin{figure}[ht]
\begin{center}
\begin{tabular}{c}
\includegraphics[height= 4.5 cm]{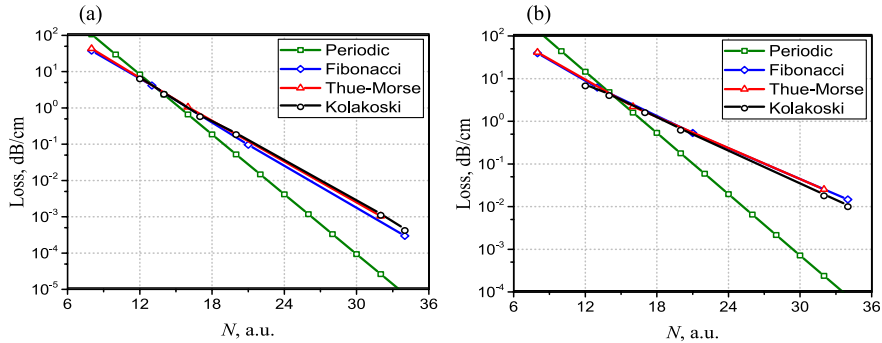}
\end{tabular}
\end{center}
\caption{Estimation of the energy leakage in the Bragg reflection waveguides  with either periodic or aperiodic configuration of layers in the finite cladding; (a) TE mode and (b) TM mode; $n_g=1$.} \label{fig:fig_2}
\end{figure}

From our estimation for the lowest modes presented in Fig.~\ref{fig:fig_2} it follows that the level of losses decreases exponentially with $N$ (see, also discussions on this matter in  \cite{west_JOSAB_2006, Argyros_OptExpress_2002}), and leakage of the TM mode is much greater than that of the TE mode. The leakage is also greater for the waveguides with an aperiodic arrangement of the cladding compared with the periodic system. Thus, the criterion of smallness of the imaginary part of $\beta$ when classifying the waveguide modes can be satisfied by setting the number $N$ to an appropriate high value.

For the symmetric waveguides all guided modes can be divided into
symmetric and anti-symmetric ones. Without loss of generality, the
overall field amplitude in the middle line of the core can be
reduced to unity or zero for symmetric and anti-symmetric modes,
respectively. Then the field in the core can be normalized by
setting $a_0=1/2$ and $b_0=\pm1/2$, where the upper sign `$+$' is
related to symmetric modes, while the lower sign `$-$' is related to
anti-symmetric ones, respectively.

\section{Numerical Results: Solution Analysis}
\label{sec:results}

Our goal here is to demonstrate a dispersion blue-shift of guided
modes in a Bragg reflection waveguide that appears in the system
having aperiodically arranged layers in the cladding. In order to
reveal the mentioned blue-shift we provide a comparative study of
the waveguide with either periodic or aperiodic configuration of the
cladding. It is supposed that the periodic Bragg mirrors consist of
a finite number of layers ($N=32$) which is enough to provide the
desired level of reflection in order to guarantee that the propagation
constant $\beta$ is essentially real. Obviously, the number of layers
of each sort $\Psi$ and $\Upsilon$ in the periodic system is the
same ($N_\Psi=N_\Upsilon=16$). Further three aperiodic designs are
investigated. In the first aperiodic configuration the mirrors are
formed by stacking together layers $\Psi$ and $\Upsilon$ according
to the Fibonacci generation rule on its eighth generation stage. At
this stage there are $N=34$ layers in the system with different
number of layers of each sort $\Psi$ and $\Upsilon$ ($N_\Psi=21$ and
$N_\Upsilon=13$). In the second configuration, the cladding is
arranged according to the Thue--Morse substitution rule on its six
generation stage which corresponds to the system with the same
number of layers as in the periodic configuration ($N=32$). There is
the same number of layers of each sort $\Psi$ and $\Upsilon$ in this
system ($N_\Psi=N_\Upsilon=16$). As the third configuration, the
cladding is considered to be an aperiodic structure arranged
according to the Kolakoski $K(1,2)$ generation scheme on its
twentieth generation stage. For such a configuration the total
number of layers coincides with those ones of the periodic and
Thue--Morse structures ($N=32$), but there is different number of
layers of each sort $\Psi$ and $\Upsilon$ ($N_\Psi=17$ and
$N_\Upsilon=15$).

For clarity, we should note that the total number of layers and the
number of layers of each sort within the deterministically aperiodic
multilayered system depend on the generation stage, and for certain
schemes it is impossible to obtain a given number of layers in the
system (e.g., using the Fibonacci substitution rule it is impossible
to obtain the system consisting of $N=32$ layers). Thereby, a
complete coincidence in the layers numbers on the earlier generation
stages exists only for periodic and Thue--Morse configurations, for
these structures the different spatial order of the layers is the
only difference between them. On the other hand, in the infinite
limit of the generation stage the numbers of layers of each sort
$N_\Psi$ and $N_\Upsilon$ in the periodic, Thue--Morse and Kolakoski
structures take on the same values, while for the Fibonacci
structure the numbers of layers can be calculated from the formulas \cite{Tuz_WRCM_09} obtained using the generating functions technique \cite{wilf_genfunct_2013}:
$N_\Psi=\left[(\tau^+)^{\sigma-2}-(\tau^-)^{\sigma-2}\right]/\sqrt{5}$,
$N_\Upsilon=\left[(\tau^+)^{\sigma-1}-(\tau^-)^{\sigma-1}\right]/\sqrt{5}$,
where $\sigma$ is the generation stage ($\sigma>2$), and  $\tau^\pm=(1\pm\sqrt 5)/2$. Here $\tau^+$ is the well known golden mean value \cite{Macia_RPP_06}.

\begin{figure}[ht]
\begin{center}
\begin{tabular}{c}
\includegraphics[height= 8.5 cm]{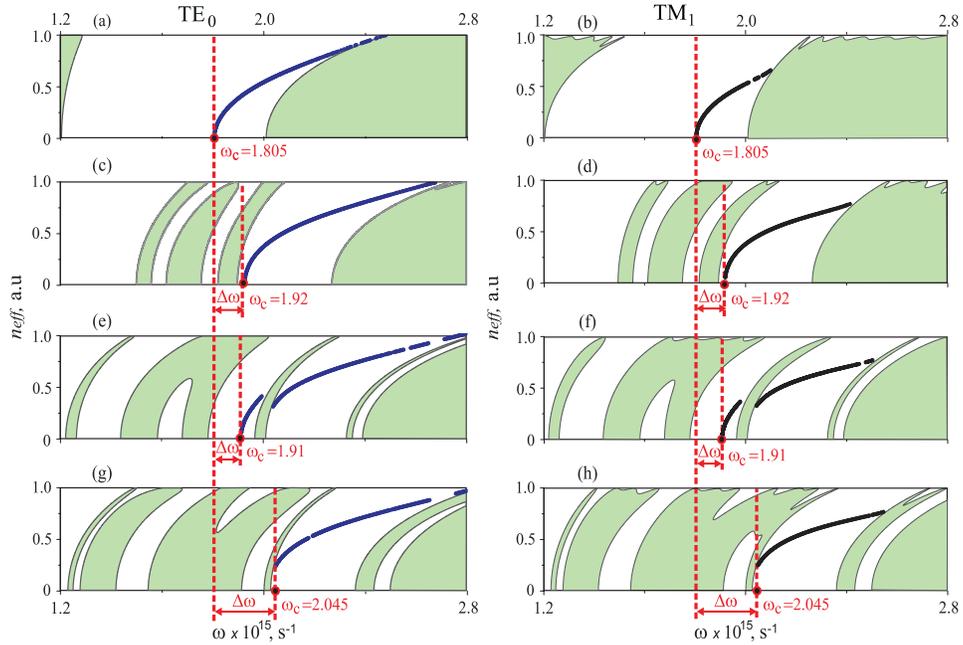}
\end{tabular}
\end{center}
\caption{The band diagrams and dispersion curves for TE (left
column) and TM (right column) polarizations in the Bragg reflection
waveguide with either periodic or aperiodic configuration of layers
in the cladding that are arranged according to different generation
rules: (a, b)~periodic; (c, d)~Fibonacci; (e, f)~Thue--Morse; (g,
h)~Kolakoski $K(1,2)$. Blue and black dispersion curves correspond
to TE$_0$ and TM$_1$ modes, respectively; $\omega_{c}$ is a cutoff
frequency; $n_g=1$.} \label{fig:fig_3}
\end{figure}

In Fig.~\ref{fig:fig_3} both band diagrams and dispersion curves for
TE and TM polarizations are presented. They are calculated for the
waveguide whose core layer is an air gap surrounded by the cladding with either periodic (a, b) or aperiodic (c-h) arrangement. Here, regions colored in light green correspond to
the bands within which light can propagate through the multilayered
structure (passbands), whereas uncolored areas correspond to the
bands where the level of reflection is high enough ($|R| > 0.99$),
therefore they can be identified as stopbands.

\begin{figure}[ht]
\begin{center}
\begin{tabular}{c}
\includegraphics[height= 4.5 cm]{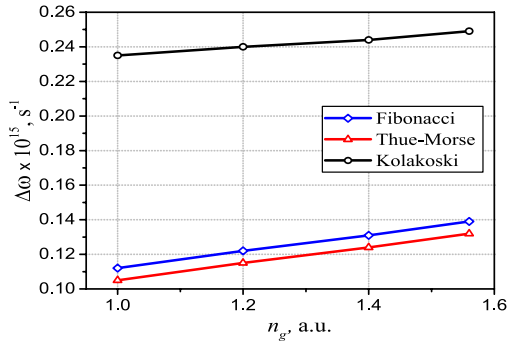}
\end{tabular}
\end{center}
\caption{The blue shift of the cutoff frequency $\omega_c$  versus the refractive index of the core layer.} \label{fig:fig_4}
\end{figure}

In this figure the guided modes are indicated by a set of colored
(blue and black) points, and it is evident that all these guided
modes appear only within the bands where the level of reflection
riches high values ($|R| \ge 0.99$) (i.e. within the stopbands). One
can conclude from Fig.~\ref{fig:fig_3}, that the dispersion curve of
each guided mode is terminated at the stopbands edge at either
maximal or minimal allowable value of $n_{eff}$, at which there is
a mode cutoff. Besides, it can be observed that for all aperiodic
waveguide configurations the dispersion curves appear to be shifted
to shorter wavelength (i.e. they acquire some blue-shift which is
marked in figures as $\Delta \omega$) as compared to the dispersion
curves related to the structure with the periodic arrangement.
 In fact this blue shift depends very little on the refractive index of
the core layer as it is depicted in Fig.~\ref{fig:fig_4} (assuming of course that $n_g$ is smaller than the
refractive index $n_\Upsilon$ of the first layer in the mirrors).

We should note that generally stopbands which carry the
above-mentioned modes are also shifted to shorter wavelength. It is
a consequence of the fact that in contrast to periodic multilayered
systems which have the main stopband centered around
$\lambda_{qw}$, aperiodic structures are characterized by a set of
(pseudo) stopbands which are distributed symmetrically around this
wavelength accompanied with highly localized transmission peaks in
the vicinity of the wavelength $\lambda_{qw}$.

Evidently, the dispersion shift appears due to different spatial
orders of the constitutive layers in the considered waveguide's
configurations. Thus, the difference in order of arrangement of
layers results in the optical thickness changing of particular
layers in the structure, since some repetitions in a row of layers
of the same sort $\Psi$ or $\Upsilon$ can appear in the aperiodic
sequences that reduces the number of layers' interfaces within the
system compared to the periodic one. The formation of layers with
doubled thicknesses $2d_\Psi$ and $2d_\Upsilon$ produces the
enlarged phase shift $\phi_i$, that leads to a change in the total
reflection spectrum of the mirror, because the shape of the
reflection spectrum of a multilayered structure depends on the phase
modulation of each layer as $\phi_i=2\pi n_i d_i/\lambda_{qw}$. If
we mark the phase modulation for the doubled layers as $\phi$,
then the reflectivity can be factorized as the product of two
contributions $R=A(n_\Upsilon/n_\Psi)\cdot \Re(\phi)$
\cite{Moretti_07}, where the function $A$ takes into account the
value of the reflectivity due to the refractive index contrast, and
$\Re$ is a shape factor that represents the different optical paths of the
light inside the layers. In our case, indeed, changes in $\Re(\phi)$
result in the dispersion shift, since the substitution
$\Re(\phi(n_i,2d_i))\to \Re(\phi(n_i,d_i))$ takes place.

It is noticed in Fig.~\ref{fig:fig_3} that the longest shift is
observed for the Bragg reflection waveguide whose cladding is formed
according to the Kolakoski $K(1,2)$ generation rule due to the greater number of the doubled layers in the mirrors compared to other considered systems.
For the given waveguide parameters, in wavelengths, the maximal shift is observed
to be about $122$~nm.

\begin{figure}[ht]
\begin{center}
\begin{tabular}{c}
\includegraphics[height= 8.0 cm]{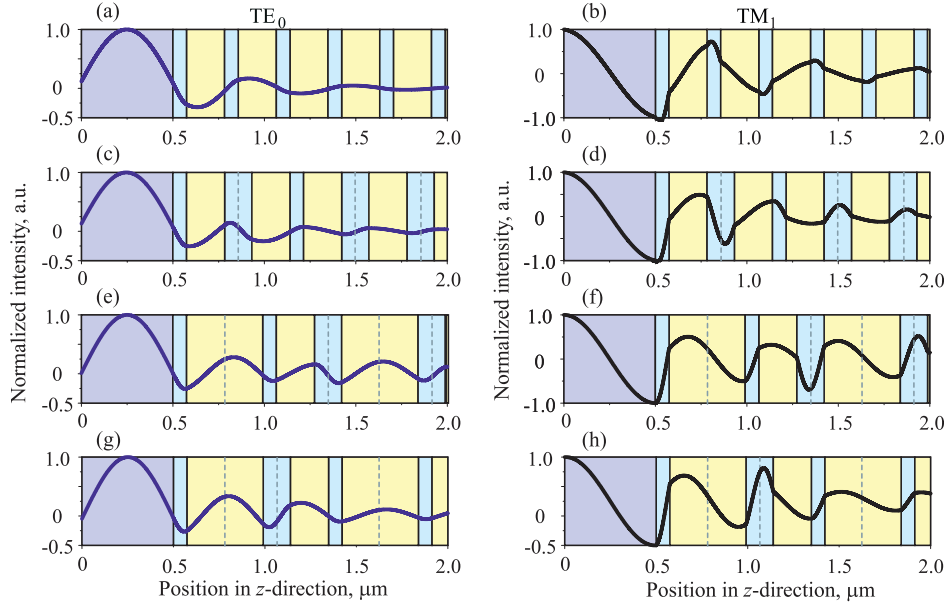}
\end{tabular}
\end{center}
\caption[example] {Intensity profile of the guided
modes for different configurations of the cladding: (a, b) periodic;
(c, d)~Fibonacci; (e, f)~Thue--Morse; (g, h)~Kolakoski $K(1,2)$.
Blue curves correspond to the electric-field patterns related to the
$E_x$ component of the field for TE$_0$ mode. Black curves
correspond to the magnetic-field patterns related to the $H_x$
component of the field for TM$_1$ mode; $n_g=1$. The confinement factor is:
(a)~$\Gamma=0.87$; (b)~$\Gamma=0.61$; (c)~$\Gamma=0.67$;
(d)~$\Gamma=0.44$; (e)~$\Gamma=0.76$; (f)~$\Gamma=0.57$;
(g)~$\Gamma=0.85$; (h)~$\Gamma=0.59$.}
\label{fig:fig_5}
\end{figure}

In order to prove that the obtained dispersion curves are associated
to the same waveguide modes, the field profiles inside the waveguide
for the corresponding field components ($E_x$ for TE mode, and $H_x$
for TM mode) are calculated at the same frequency ($\omega=2.1 \times
10^{15}$~s$^{-1}$) which are reported in Fig.~\ref{fig:fig_5}. These
mode-field profiles are plotted after applying normalization of the
field magnitude on its maximal value within the core. From these
figures it follows, that the field of the fundamental TE$_0$ mode
appears to be well confined within the central air-guiding core. The
degree of this confinement can be estimated via calculation of the
confinement factor
$\Gamma=(\int_{-d_g}^{d_g}|I|^2\,dz)/(\int_{-\infty}^\infty
|I|^2\,dz)$,  where $I$ is related to field components $E_x$ and
$H_x$ for TE modes and TM modes, respectively. Thus, the
confinement factor related to the TE$_0$ mode manifests rather
high values for all discussed configurations (see, values of
$\Gamma$ given in the caption of Fig.~\ref{fig:fig_3}) and, especially,
the highest values $0.87$ and $0.85$ are obtained for the periodic and
aperiodic Kolakoski $K(1,2)$ configurations of the cladding.
Besides, the field intensity of the fundamental mode decays
rapidly within the first few pairs of the constitutive layers of the
cladding as it is depicted in Fig.~\ref{fig:fig_3}. At the same
time, the field of TM$_1$ mode appears to be less confined within
the core layer, thus, as was already mentioned, the field leakage is higher
for this polarization.

\section{Conclusions}

To conclude, a dispersion blue-shift in a Bragg reflection waveguide
consisting of an aperiodic design of layers in the cladding compared
to the periodic design is demonstrated. The strength of this
blue-shift is investigated for three aperiodic multilayered systems,
namely for those alternated according to Fibonacci, Thue--Morse and
Kolakoski substitution rules. It is recognized that the longest
shift is observed for the Bragg reflection waveguide whose cladding
is formed according to the Kolakoski $K(1,2)$ sequence.

In order to prove that the considered dispersion curves are
associated to the same waveguide modes, both the field profile
inside the waveguide and the confinement factor for each mode are calculated. It is
found out that the highest confinement factor can be achieved in the
Kolakoski $K(1,2)$ structure among other aperiodic designs.

We argue that the design of cladding in the form of an aperiodic
structure gives rise to change in the dispersion characteristics,
resulting in the shift of cutoff wavelengths of guided modes toward
the shorter wavelength regardless of their polarization.

\appendix
 \section{Aperiodic Orders}
\label{sec:ApA}

In optics, a standard algorithm for arranging aperiodic photonic
structures is usually based on a certain symbolic substitution rule
(i.e. on a specific substitution rule $w$ that operates on a finite
alphabet $A$ which consists of a set of letters $\{a,b,c,...\}$). In
practical realizations each letter can be associated with a
particular constitutive block (e.g. with a dielectric or
semiconductor layer) in a resulting photonic structure. In this
regard, substitution sequences that act upon a two-letter alphabet
are especially important. Such algorithms can be reduced to
following:
\begin{equation}
    \label{eq:substitutionalsequences}
    a \to w_1(a,b),~~~b \to w_2(a,b),
    \end{equation}
where $w_1$ and $w_2$ can be any string of letters $a$ and $b$.

Fibonacci sequence is obtained via iteration of the rule $w_F$
\cite{Kohmoto_PhysRevLett_1987}, as
\begin{equation}
    \label{eq:Fib}
    a \to ab,~~~b \to a,
\end{equation}
resulting in the letters chain $\{a \to ab \to aba \to abaab \to
abaababa \to ...\}$.

Thue--Morse sequence is generated by the substitution $w_{TM}$
\cite{Liu_PhysRevB_1997}, as
\begin{equation}
    \label{eq:TM}
    a \to ab,~~~b \to ba,
\end{equation}
resulting in the letters chain $\{a \to ab \to abba \to abbabaab \to
...\}$.

The generation rule $w_K$ of the Kolakoski sequence is similar to
those ones of Fibonacci and Thue--Morse sequences and can be based
on two symbols substitution. Namely, the sequence $K(a, b)$ can be
obtained by starting with $a$ as a seed and iterating the following
two substitutions \cite{Sing04}
\begin{equation}
    \label{eq:Kpq}
    w_0~:
    \begin{array}{ccccccccccccc}
          b~\to a^{b}\\
          a~\to a^{a}\\
    \end{array}
~~~~~~~\mbox{and}~~~~~~~~w_1:
    \begin{array}{ccccccccccccc}
          b~\to b^{b}\\
          a~\to b^{a}\\
    \end{array}
\end{equation}
where $w_0$ and $w_1$ can be any string of letters $a$ and $b$;
$a^b$ denotes a run of $a$ $b$'s, i.e., $a^b = a...a$ ($b$ times).
In this paper we examine the case when $a=1$ and $b=2$, so the
letters chain $\{abbaababba...\}$ appears. Remarkably, using the
Kolakoski substitutional rule (\ref{eq:Kpq}) it is possible to obtain
any value of the total number of letters in the resulting chain by
changing its generation stage and starting letter of the sequence.

\section{Transfer Matrix Description}
\label{sec:ApB}

 The plane monochromatic waves of TE ($\vec
E^{\mathrm{TE}} \parallel \vec x_0 $) and TM ($\vec H^{\mathrm{TM}}
\parallel \vec x_0 $) polarizations can be defined in a particular
$j$-th layer ($j=1,2,...,N)$ of the sequence in the form
\begin{equation}
\label{eq:fieldcomponents} \left\{ \begin{matrix} \vec
E^{\mathrm{TE}}_j \\ \vec H^{\mathrm{TM}}_j
\end{matrix}\right\}=\vec x_0\left\{ \begin{matrix}
1/\sqrt{Y_j^{\mathrm{TE}}} \\ \sqrt{Y_j^{\mathrm{TM}}}
\end{matrix}\right\}u_j(z)\exp[-i(\omega t - \beta y)],
\end{equation}
where
\begin{equation}
\label{eq:amplitude} u_j(z)=a_j\exp(ik_{zj}z)+b_j\exp(-ik_{zj}z),
\end{equation}
and $a_j$ and $b_j$ are the field amplitudes,
$Y_j^{\mathrm{TE}}=k_{zj}/k_0\mu_j$ and
$Y_j^{\mathrm{TM}}=k_0\varepsilon_j/k_{zj}$ are the wave
admittances,  $k_0=\omega/c$ is the wavenumber in free space, and
$k_{zj}$ is the transverse wavenumber which takes on discrete values
in each slab and can be written as
\begin{equation}
\label{eq:wavenumber} k_{zj}=\left(k_0^2n_j^2
-\beta^2\right)^{1/2}=k_0\left(n_j^2-n_{eff}^2\right)^{1/2}.
\end{equation}
Here $n_j$ takes on values $n_g$ for the core layer, and $n_\Psi$
and $n_\Upsilon$ for the $\Psi$ and $\Upsilon$ cladding layers,
respectively.

The field amplitudes for the structure input and output are
evaluated as \cite{Tuz_JOSA_09, Tuz_WRCM_09, Tuz_JModOpt_10}
\begin{equation}
\label{eq:matr1}
\begin{array}{ccccccccccccc}
\left[ \begin{matrix}a_{0}\\b_{0}\end{matrix}\right] =
\mathbf{T}_\Sigma\left[
\begin{matrix}a_{N}\\b_{N}\end{matrix}\right] = &
(~\underbrace{\mathbf{T}_\Psi \mathbf{T}_\Upsilon
\mathbf{T}_\Upsilon \mathbf{T}_\Psi \mathbf{T}_\Psi
\mathbf{T}_\Upsilon\ldots }~)& \left[
\begin{matrix}a_{N}\\b_{N}\end{matrix}\right],
\\
           & N &
\end{array}
      \end{equation}
where the total transfer matrix $\mathbf{T}_\Sigma$ is obtained by
multiplying in the appropriate order the matrices corresponding to
each layer in the structure.

The matrices $\mathbf{T}_\Psi$ and $\mathbf{T}_\Upsilon$ are the
particular transfer matrices of rank 2 of the $\Psi$ and $\Upsilon$
layers in cladding with their corresponding thicknesses $d_\Psi$ and
$d_\Upsilon$. They are
\begin{equation}
\label{eq:transferUP}
\mathbf{T}_\Psi=\mathbf{T}_{01}\mathbf{P}_1(d_\Psi)\mathbf{T}_{10},~~~
\mathbf{T}_\Upsilon=\mathbf{T}_{02}\mathbf{P}_2(d_\Upsilon)\mathbf{T}_{20},
\end{equation}
where $\mathbf{T}_{0j}$ and $\mathbf{T}_{j0}$ ($j = 1, 2$) are the
transfer matrices of the layer interfaces with outer half-spaces,
and $\mathbf{P}_j(d)$ are the propagation matrices through the
corresponding layer. The elements of the matrices $\mathbf{T}_{0j}$
and $\mathbf{T}_{j0}$ are determined by solving the boundary-value
problem related to the field components (\ref{eq:fieldcomponents}):
\begin{equation}
\label{eq:matr01}
\mathbf{T}_{pv} = \frac{1}{2\sqrt{Y_pY_v}}\left[
\begin{matrix}
Y_p+Y_v & \pm( Y_p-Y_v) \\
\pm( Y_p-Y_v) & Y_p+Y_v
\end{matrix}
\right],
\end{equation}
\begin{equation}
\label{eq:matr02}
\mathbf{P}_{j}(d) = \left[
\begin{matrix}
\exp(-ik_{zj}d) & 0 \\
0 & \exp(ik_{zj}d)
\end{matrix}
\right],
\end{equation}
where the upper sign `$+$' relates to the TE wave, while the lower
sign `$-$' relates to the TM wave.

Finally, the reflection coefficient of the layer stack is determined
by the expression
\begin{equation}
\label{eq:Ref}
 R=|R|\exp(i\phi) = (b_0/a_0)|_{b_N=0}=-t_{21}/t_{22},
 \end{equation}
where $t_{mn}$ are the elements of the transfer matrix
$\mathbf{T}_\Sigma$, and $\phi$ is the phase of the reflected light.


\bibliographystyle{elsarticle-num}
\bibliography{FesenkoTuz}






\end{document}